\documentclass[12pt]{iopart}
\usepackage{iopams,epsfig}

\def\fullsquare{\vrule height 4pt depth -1pt width 3pt\kern 0.1em}
\newcommand{\mum}{$\mathrm{\,\mu m}$}

\begin{document}

\title{Tunable whispering gallery modes for spectroscopy and CQED experiments}

\author{Wolf~von~Klitzing\dag $\|$, Romain~Long\dag, Vladimir~S~Ilchenko\S, Jean~Hare\dag,
Val\'erie~Lef\`evre-Seguin\dag}
\address{\dag~Laboratoire Kastler Brossel, D\'epartement de Physique de l'Ecole Normale Sup\'erieure, 24 Rue Lhomond, 75231 Paris Cedex 05, France}
\address{$\|$~Now at the Quantum Gases group of the FOM Institute for Atomic and Molecular Physics, Kruislaan 407, 1098 SJ Amsterdam, The Netherlands}
\address{\S~Now at Jet Propulsion Laboratory, California Institute of Technology, 4800 Oak Grove Drive, Pasadena, California 91109-8099\\email: Wolf.vonKlitzing@lkb.ens.fr, Valerie.Lefevre@lkb.ens.fr}


\begin{abstract}
We have tuned the whispering gallery modes of a fused silica
micro-resonator over nearly 1\,nm at 800\,nm, i.e.~over half of a
free spectral range or the equivalent of $10^6$ linewidths of the
resonator. This has been achieved by a new method based on the
stretching of a two-stem microsphere. The devices described below
will permit new Cavity-QED experiments with this ultra high
finesse optical resonator when it is desirable to optimise its
coupling to emitters with given transition frequencies. The tuning
capability demonstrated is compatible with both UHV and low
temperature operation, which should be useful for future
experiments with laser cooled atoms or single quantum dots. A
general overview of the current state of the art in microspheres
is given as well as a more general introduction.
\end{abstract}


\maketitle


\section{Introduction}

Optical micro-cavities have attracted much interest in the field
of quantum cavity
electrodynamics\cite{CampilloEversole91,TreussartHaroche1994} as
well as in classical and nonlinear
optics\cite{SchillerByer91,IlchenkoGorodetsky92,LinCampillo94}.

Very low mode-volume semiconductor cavities using multi-layer
dielectric mirrors have been developed. Low threshold lasing
\cite{HeitmannYamamoto1993,KlitzHaroche99} and cavity enhanced
spontaneous emission \cite{DeMartiniInnocenti87,Eta_Yamamoto92}
have been observed. Small Fabry Perot cavities have a larger
mode-volume but can achieve very high quality factors
($Q=\nu/\Delta\nu$). The strong coupling regime has been reached
between single alkali atoms and the fundamental mode of the cavity
\cite{PinkseRempe00,HoodKimble2000}. The resulting Rabi-splitting
of the coupled mode \cite{ThompsonRempe92}, as well as the single
atom laser action \cite{AnChilds94} have been demonstrated.  More
recently an atom has been trapped by a single photon in such a
cavity and its motion deduced from the optical
signature.\cite{HoodKimble2000,PinkseRempe2000}

An attractive alternative to Fabry Perot cavities are solid
dielectric microspheres having at the same time a very low
modevolume \emph{and} very high quality factors. Light can be
trapped in so called whispering gallery modes (WGMs) if the
refractive index of the material of the sphere is larger than the
one surrounding it.\footnote{WGMs had first been observed in the
gallery of the cupola of St.Paul's Cathedral in London. A whisper
spoken close to the wall can be heard all the way along the
gallery, some 42\,m to the other side. Lord Rayleigh was the
first to identify the refocusing effect of the curved surface as
the sound travels along the gallery.  He also conjectured the
existence of the thus called whispering gallery modes. He also
suggested that such modes of the electromagnetic field could find
some applications due to the extreme confinement of the
field.\cite{Logan65}} Successive total internal reflections off
the concave inner surface confine the light into a thin ring
close to the equator. These high-$Q$ ring modes have been
observed and studied extensively in
droplets.\cite{ChangCampillo96} It was soon recognised that the
field enhancement caused by the strong confinement of the light in
these modes combined with their high quality factors could lead to
strong Cavity QED effects. Stimulated Raman scattering, to name
but one, has been observed in small CS${_2}$ droplets with a
threshold of just three photons per mode.\cite{LinCampillo94}
Whispering-gallery modes have also been used to produce lasers in
microdisks\cite{McCallLevi92} and to enhance the spontaneous
emission of quantum dots in micropillar stuctures\cite{Gerard98}.

However, semiconductor microdisks and pillars can only operate in
the low-$Q$ regime and microdroplets suffer unfortunately from
evaporation and gravitational pull. This problem has been overcome
in the pioneering work of V~Braginsky et al.
\cite{BraginskyGorodetsky89}. He and his coworkers realised that
whispering-gallery modes in silica micropsheres unique combination
of small mode resonators with ultra high $Q$ factors. Moreover,
these spheres remain attached to a thin stem of silica and thus
are easily manipulated and are easily produced in the laboratory.
For microspheres ($\varnothing\simeq40$\mum) the modevolume can be
exceedingly small ($V\sim100$\mum$^3$). The electrical field for a
\emph{single photon} in such a mode is of the order of 10\,kV/m.
At the same time the quality factor of, e.g. silica microspheres,
can be as high as $10^{10}$ with photon storage times of the order
of one microsecond.\cite{CollotLefevre93,GorodetskySavchenkov96}
Clearly such a system is ideally suited for the observation of
non­linear optical effects and Cavity­QED experiments.



\section{WHISPERING GALLERY MODES IN MICROSPHERES}

This section gives a short overview over the general properties of
whispering gallery modes (WGMs). A more detailed account of the
theory of the WGMs and on experiments performed on microspheres
can be found in References~\cite{BarberChang88,ChangCampillo96}.

\begin{figure}
  \begin{center}
  \psfig{figure=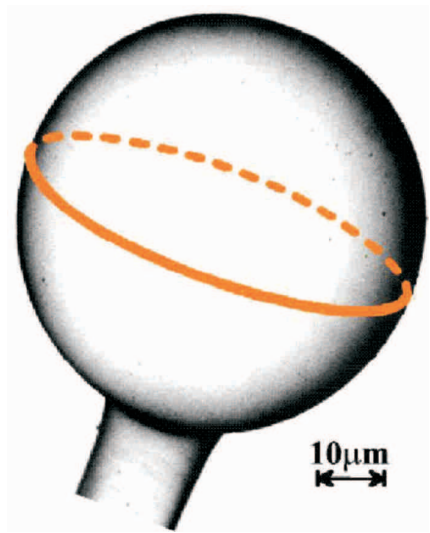,height=7cm}
  \end{center}
  \caption{Scanning electronic microscope image of a fused-silica microsphere, 56 microns in diameter. }
  \label{fig:sphere}
\end{figure}

A transparent dielectric sphere can sustain WGMs if its
circumference is larger than a few wavelengths.
Figure~\ref{fig:sphere}) shows a typical single-stemmed silica
microsphere produced in our laboratory. The WGMs can be understood
as high angular momentum electromagnetic modes in which light
propagates by repeated total internal reflection surface at
grazing incidence with the proper phase matching condition. The
modes can readily be derived from Maxwell's equations solved in
spherical coordinates. The angular dependence of the field is
naturally described with spherical harmonics. The two quantum
numbers $l$ and $m$ (with $m=-l,\ldots+l$) describe the total
angular momentum and its projection upon the reference axes
respectively. Quantum numbers $m$ of opposite sign correspond to
waves propagating in opposite directions along the perimeter of
the sphere. The modes offering the highest polar confinement and
thus the smallest modevolume correspond to values $|m|$ close
to~$l$. In the radial direction the index discontinuity creates a
potential well which combines with the centrifugal barrier to form
a pocket like pseudo-potential. This effective potential approach
has been analyzed in detail by H M Nussenzveig
\cite{GuimaraesNussenzveig94}. It provides good physical insight
in many properties of the WGMs which appear as quasi­bound states
of light, analogous to the circular Rydberg states of alkali
atoms. The radial confinement of the WGMs is characterized by the
quantum number $n$, the number of antinodes of the field
amplitude. The modes with $n=1$ are the most confined in radial
direction both in terms of the modevolume and `leakage'. In the
image of geometrical optics these modes undergo a maximum of
reflections at the surface and thus have the smallest reflection
angle, the lowest diffraction losses. Modes near this condition
have a free spectral range FSR\,$\cong c/(N l \lambda)$. The
boundary conditions imposed on the field depend on its
polarization: Modes with TE and TM polarization will undergo
different phase-shifts upon reflection at the surface of the
sphere. Two modes which differ only in polarisation will exhibit
resonance frequencies a substantial fraction of an FSR apart.

In short, once the polarization of the mode is assigned, WGM's are
described by three integers $n$, $l$, and $m$. The mode with the
longest life time and the smallest volume is $(l=l_{max}, |m|=\pm
l, n=1)$. It is confined near the bottom of the potential well,
i.e. as close as possible to the sphere's surface. It has a
maximum angular momentum of $l\simeq Nka\simeq Nx$, with $x$ being
the size parameter ($x=ka= 2\pi a/\lambda $) of a sphere of radius
$a$. Its cross section is almost Gaussian both in polar and radial
direction.

Light trapped in these modes can escape out of the sphere only by
tunnelling across the potential barrier which extends as far as
$Na$ for this state. The very short evanescent tail
($\simeq\lambda/2\pi$) of the quasi-bound state $(n\cong1)$
implies a very weak coupling to the outside medium and thus
extremely high quality factors $Q=\nu/\Delta\nu$. In highly
transparent media such as fused synthetic silica the diffraction
losses are negligible for spheres larger than $a=20$\mum. A given
$Q$ value is related to an attenuation $\alpha$ by the formula $Q
= 2\pi N/(0.23\alpha \lambda$) where $\alpha$ is expressed in
dB/m. The minimum attenuation observed in this wavelength region
on silica optical fibres is about 3\,dB/km corresponding to
$Q=2\times10^{10}$. Absorption, scattering on impurities, and
residual surface roughness limit in praxis the $Q$ to about
$3\times10^9$ at
780\,nm\cite{BraginskyGorodetsky89,CollotLefevre93,GorodetskySavchenkov96}
corresponding to an attenuation of 17\,dB/km.

\subsection{Coupling light into whispering gallery modes}

\begin{figure}
  \begin{center}
  \psfig{figure=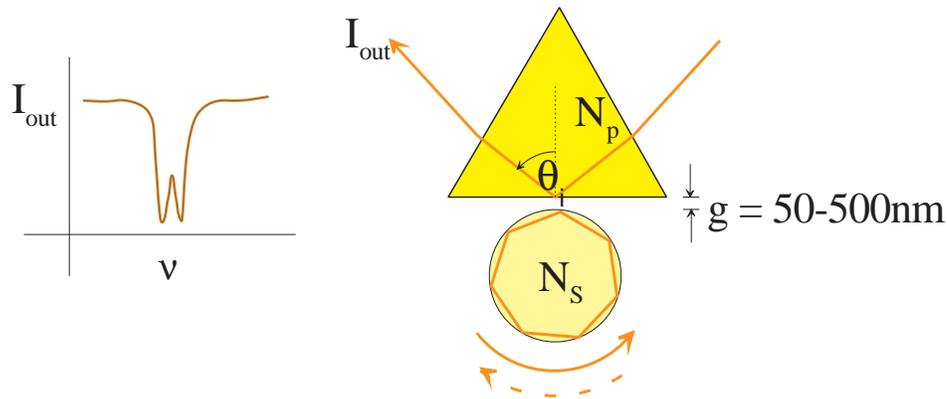,height=5.2cm} 
  \end{center}
  \caption{Approaching a high refractive index prism to the sphere (not to scale) one can frustrate the total internal reflection of the WGMs within the sphere and couple light into or out of the sphere.}
  \label{fig:schem-setup}
\end{figure}

The very high diffraction limited $Q$ of the fundamental modes of
larger spheres implies that free space coupling to the
microspheres is very inefficient. In order to achieve efficient
coupling the free space beam has to be matched to the WGM. Close
to the surface of the sphere the excitation beam has to have the
same shape and angular momentum (with respect to the centre of the
sphere) as the mode. The potential barrier created by the index
discontinuity at the surface of the sphere confines almost all of
the field to the inside of the sphere. However, a short evanescent
tail of the electro magnetic field protrudes from the sphere. If a
material of high refractive index is brought into this evanescent
wave some of the light will tunnel across the gap between the
material and the sphere, also known as frustrated total internal
reflection. This can be achieved with a prism of high refractive
index $(Np)$ almost in contact with the sphere. If an incident
beam hits the prism surface with an angle $\theta$ close to the
critical angle $\theta=\arcsin (N/ Np)$ so that its angular
momentum with respect to the centre of the sphere is
$N_{p}ka\sin\theta \simeq Nka$ its light can be fed into a WGM. By
slightly changing the angle of incidence and the frequency of the
beam different WGMs can be selectively excited.\footnote{This
method has also been used in the pioneering work of Braginsky et
al.\cite{BraginskyGorodetsky89} and in another geometry in
Paris\cite{CollotLefevre93} where also eroded fibre couplers and
tapered fibres have been used \cite{DubreuilKnight95}.} Using the
prism coupling scheme efficiencies exceeding 30\% can be achieved
for various mode geometries. The resonances appear as dips in the
intensity of the beam reflected from the prism (figure
\ref{fig:schem-setup}). The depth of these dips is a direct
measure of the coupling efficiency. Due to the nature of
evanescent waves the coupling rate decreases exponentially with
increasing gap size. The ability to control the coupling rates is
an important advantage of microspheres as compared to other
resonators where the coupling rate is fixed by the reflectivity of
the mirrors. The highest coupling efficiency between the free
space modes and the WGMs is achieved with a gap of a few hundred
nanometres where the coupling rate matches the other losses of the
resonator (generally absorption and diffraction by surface the
roughness). The intrinsic quality factors of up to $10^{10}$ can
only be measured with very large gaps (typically of the order of
1\mum).

\begin{figure}
  \begin{center}
  \psfig{figure=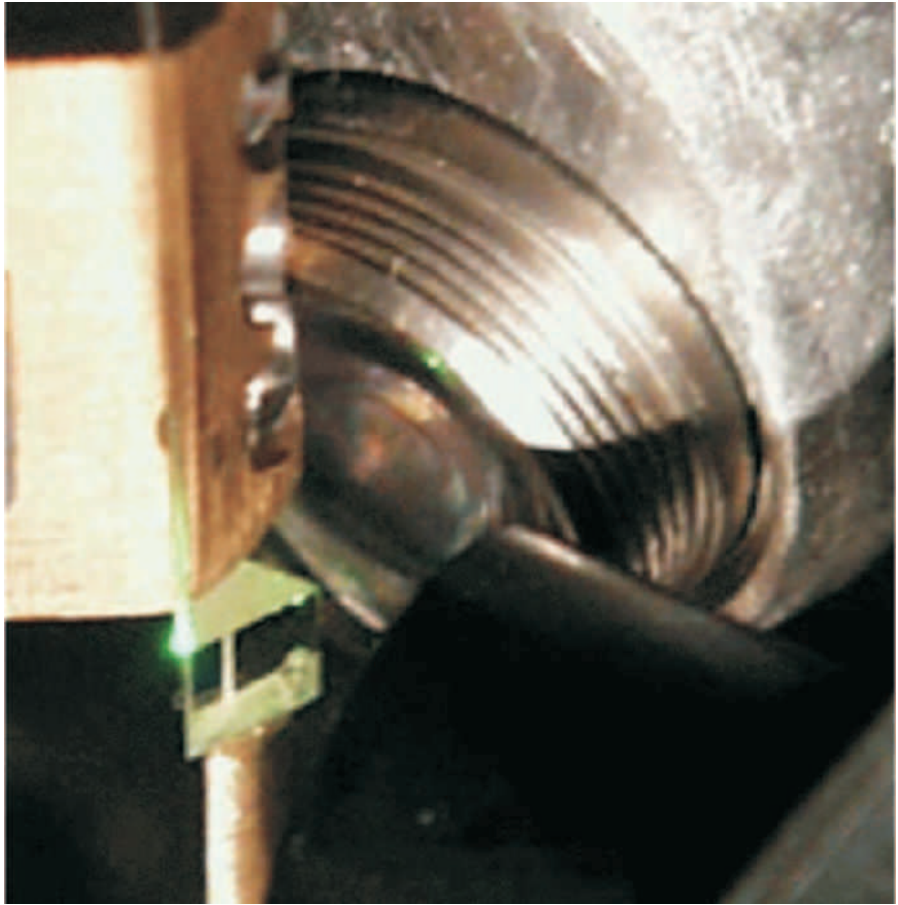,height=12cm} 
  \end{center}
\caption{A photo of the central part of the experiment. The brass
holder (top left of the picture) holds a single stemmed Er$^{3+}$
doped ZBLAN sphere ($\varnothing=140$\mum). An aspherical lens
(lower right side) projects the light (10\,mW, 800\,nm) from an
optical fibre onto the prism. Part of the laser light is absorbed
by the microsphere and re-emitted at 550\,nm. Some of the
fluorescent light exiting the sphere can be seen in the photo as a
bright spot near the corner of the prism. The lens (top right
side) collimates the remaining IR pump radiation and the emitted
green laser light. It can thus be analysed outside of the
temperature controlled central part of the experiment.}
  \label{fig:erbium}
\end{figure}

\section{Experiments on microspheres in Paris}

\subsection{Experimental set up}
\begin{figure}
  \begin{center}
  \psfig{figure=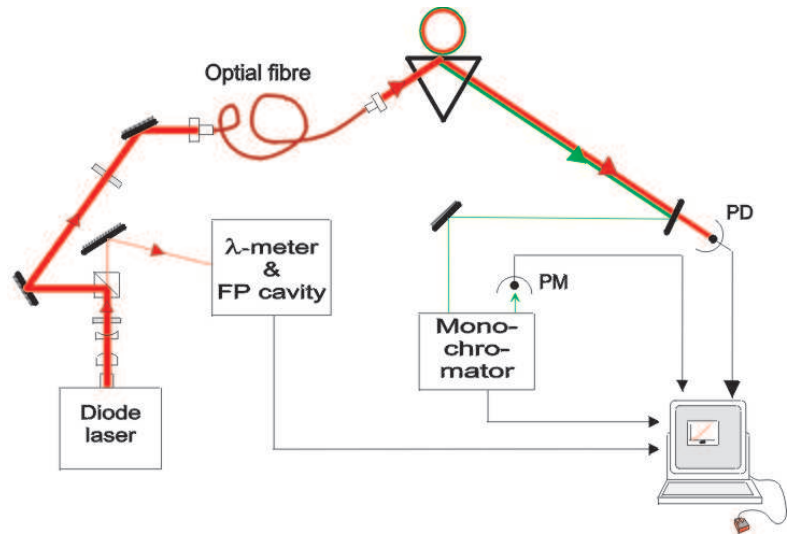,height=7cm}
  \end{center}
  \caption{Approaching a high refractive index prism to the sphere (not to scale) one can frustrate the total internal reflection of the WGMs within the sphere and couple light into or out of the sphere.}
  \label{fig:schem-setup-all}
\end{figure}

The experiments described here are all based on the prism coupler
method: The light is coupled into and out of the sphere via a
prism. As described above the incoming beam has to match the
whispering gallery mode in frequency, spot size, and angular
momentum. It is therefore necessary to have full control over all
parameters of the incoming Gaussian beam: frequency, waist
diameter and position, and angles of incidence with the prism.
Figure~\ref{fig:schem-setup-all} shows a schematic of the setup
used in the experiments presented here. A photographic image of
the coupling lenses and the prism can be seen in figure
\ref{fig:erbium}. When narrow linewidth is required, e.g. for a
measurement of the ultimate $Q$ a grating stabilised laser was
used. A laser diode (Yokogawa YL78XNL) with integrated Bragg
grating serves when a large tuning range is needed. A wavemeter
laser provides the absolute frequency scale with a precision of
about $10^{-4}$\,nm. A small fraction of the laser light
($\sim3\%$) traverses a calibrated high-finesse Fabry Perot
cavity. The transmission peaks of the Fabry Perot cavity are
recorded together with any absorption data taken and thus give a
precise relative frequency scale. The remainder of the light
couples into a single mode fibre after intensity and polarization
control. This fibre leads into a temperature controlled box
containing the launching and collection optics as well as the
microsphere with the necessary mechanical controls.  A high
aperture lens collimates the light exiting the fibre onto the
equilateral coupling prism (SF11). The angles of incidence and
the numerical aperture are calculated in advance as a function of
the size of the sphere and accordingly adjusted before
introducing the microsphere. The light coupled out of the sphere
is collimated and analyzed outside the confinement. The
microsphere itself is mounted on 3D micrometer translation
stages. The gap between the sphere and the prism can be adjusted
with nanometric precision using a low voltage piezostack.

\subsection{Tuning Microspheres}
Until recently the main default of the solid microspheres has been
that the frequencies of the resonances are not tunable. The free
spectral range (FSR) in small spheres is of the order of THz
whilst the line width of the resonances is about 300\,kHz.
Therefore accidental coincidences between an atomic line and the
fundamental transverse whispering gallery mode are extremely rare.
We have recently demonstrated a tuning device capable of spanning
up to an FSR. This opens a whole new range of experiments to solid
dielectric microspheres. Now fixed dipoles can be brought into
interaction with these resonators. Applications include CQED
experiments with cooled atoms, or cavity-ring-down spectroscopy of
environmentally important gases. Recently the first practical
tuning device has been developed in our
laboratory.\cite{vonKlitzingLefevre2000}.

A number of conditions have to be fulfilled in order for the
tuning device to be useful: First of all it must safeguard the
high quality factor and function for small spheres to take
advantage of the reduced mode volume. The tuning range should be
of the same order of magnitude as the free spectral range (FSR)in
order to be able to tune a desired WGM into resonance with, e.g.,
an atomic transition or resonances in quantum dots. The device has
to be exceedingly stable. A change by only $10^{-7}$ of the
desired tuning range would already shift the WGMs by one resonant
linewidth. Good access to the sphere must be safeguarded in order
to be able to approach a prism from one side and the sample the
sphere is to interact with from the other. Furthermore the device
should be readily producible and affordable, which is especially
important for potential applications, e.g. as trace gas detectors.
In many cases vacuum compatibility and/or low temperature
operation would be highly desirable. Here two devices are rapidly
presented which fulfils all these conditions with the first one
aiming more for spectroscopic applications the other for the more
demanding CQED experiments. Some detail as to the production are
going to be given and some examples of level (anti-) crossings
will be presented.

In principal there are two methods to tune whispering gallery
modes in solid dielectric microspheres: temperature
\cite{VernooyKimble1998} and strain \cite{IlchenkoHaroche1998}. At
first order, both affect the mode resonance through the simple
relation: $\Delta \nu/\nu = -\Delta a/a -\Delta N/N$, where $a$ is
the radius of the sphere and $N$ its refractive index. The
temperature dependence of the modes is about $-2.5$~GHz/K. Given a
free spectral range of about 1\,THz for a microsphere of 60\mum\
diameter this can only serve as fine tuning. On the other hand
silica glasses can be deformed elastically up to a few percent.
One FSR tuning is equivalent to $\Delta \nu/\nu \simeq 1/l$ where
$l$ is the longitudinal quantum number. For a typical sphere this
implies an equatorial deformation of 0.2\% are sufficient, which
can be achieved in silica.\cite{HustonEversole93} The first
demonstration of strain tuning used piezodriven pliers in order to
\emph{compress} the microsphere.\cite{IlchenkoHaroche1998} About
one quarter of the sphere protruded from the device thus allowing
coupling to the WGMs. For a sphere of a diameter of 160\mum\
tuning over 150\,GHz at 800\,nm has been demonstrated. However,
the jaws restrict the access to the sphere and the device can not
be applied to spheres smaller than about 100\mum. This precludes
its use in experiments on, e.g., quantum dots (access) and
thresholdless lasing in Nd doped silica (size).

A new method has been developed recently in our laboratory in
which the strain is applied to the sphere by \emph{stretching} it.
\cite{vonKlitzingLefevre2000} We are now able to produce spheres
with two stems, one on each pole. The strain on the microsphere
can therefore now be exerted simply by pulling on the ends of the
two stems. The requirements on the symmetry are rather stringent
though: upon pulling the two stems even a slight angle between the
two results in very large forces between the sphere and the stems
and thus to early breakage.


\subsection{The Production of tunable silica microspheres}

The starting material for the microspheres is a rod of synthetic
silica glass.  This is being heated in an oxygen-propane torch and
rapidly pulled into a fibre of 20--60\,\mum\ diameter. The sphere
is created again by heating part of the fibre upon which the
surface tension pulls the molten material into an approximately
spherical shape. An industrial 10\,W $\mathrm{CO_2}$-laser serves
as a highly controllable, clean source of heat. The process is
controlled by eye through a binocular microscope. The focusing
lenses and the glass fibre itself are mounted on micrometer
controlled 3D translation stages.

A lens (f\,=\,25\,mm) focuses the vertical 3\,mm diameter
$\mathrm{CO_2}$-laser beam to a waist of about 30\mum. It thus
creates a strong vertical gradient in the intensity of the
infrared radiation. The fibre hanging down is only heated near the
focus. The glow of the silica due to the heating is visible in a
stereo microscope and serves as an indication of the temperature
of the glass. (Grey and UV filters have to be used to protect the
eyes.) Close to the melting point the surface tension starts to
pull the glass into a round shape. However, since the laser light
comes from below this shape will cast a shadow upon itself and
little radiation reaches the equator. As a consequence not a
sphere but a pear shaped object is being formed. This can be
corrected for by slowly moving the sphere down past the free space
focus of the $\mathrm{CO_2}$-laser towards the point of maximum
wave front curvature.  The shadow of the sphere now lies on the
stem above the sphere and protects it from the heat whilst the
large aperture of the incident light allows the equator
efficiently to be heated. As can be seen on the left hand side of
Fig.\ref{fig:device-v1} the result is rather symmetrical with
respect to the equatorial plane. The production of such a sphere
including the fibre preparation takes only about ten to twenty
minutes on a well adjusted laser set up.

As mentioned above it is absolutely crucial to preserve the
symmetry of revolution of the microsphere and stems. Otherwise the
stems will rupture prematurely. Our production setup obeys this
symmetry: The $\mathrm{CO_2}$-laser beam and the fibre both are
strictly vertical.  However, once the melting process has started
the slightest current of air will move the lower stem to one side.
In order to guarantee the fibre being vertical we thus use a small
weight (5--10\,mg) attached to its lower end. Additionally before
producing the actual sphere residual tension in the fibre, caused
e.g. by imperfect mounting, is removed by annealing it with the
$\mathrm{CO_2}$-laser: Close to but below the melting point the
residual stress relaxes on a time scale of a few seconds.


\subsection{The tuning device \#1}
\begin{figure}
  \begin{center}
  \psfig{figure=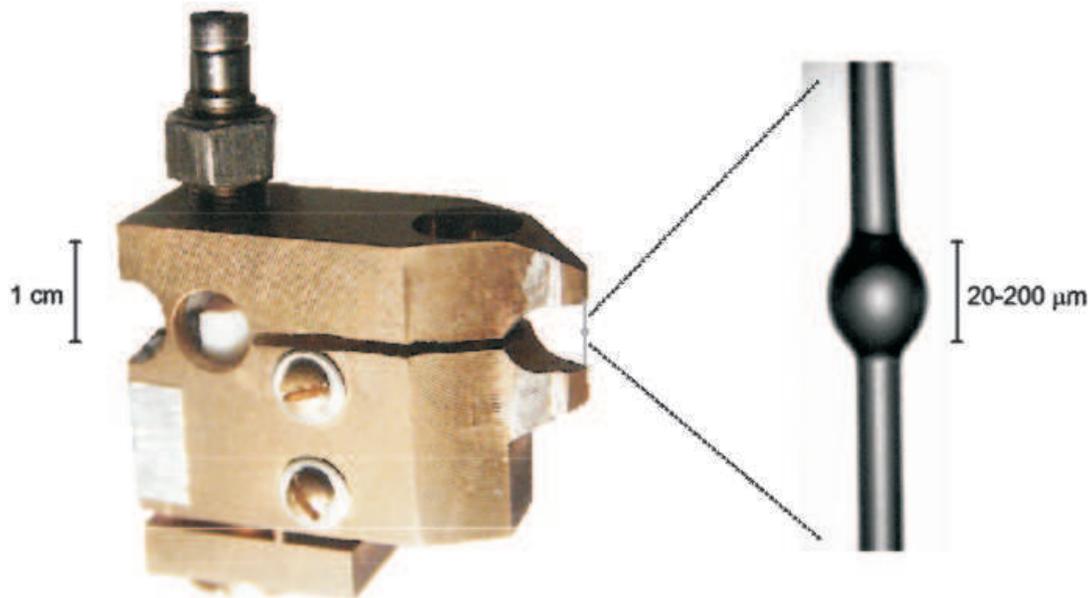,height=8cm}
  \end{center}
  \caption{The first tuning device. Two brass jaws hold a double
stemmed sphere of a diameter between 60 and 200\mum. The WGMs of
the sphere can be tuned by stretching it using a fine screw and a
low voltage PZT stack. The right hand side of the figure shows a
CCD camera microscope image of a typical microsphere.}
  \label{fig:device-v1}
\end{figure}

Figure \ref{fig:device-v1} shows the tuning device \#1. The sphere
with its two stems is glued between two brass arms which can be
opened and closed with fine screws and a low voltage PZT-stack. At
the tips of the two arms are U shaped notches of the dimensions
$0.1\times0.5\times5\,$mm with a 3--5\,mm gap between them. Into
these the stems of the sphere are fixed using standard
cyanoacrylate glue. The set screw is then tightened to remove the
inevitable slack in the stems so that the PZT can exert strain on
the stems and thus the sphere. Spheres with a diameter down to
about 60\mum\ and a stem diameter of about 40\mum\ can be easily
used in this device. Much thinner stems break too easily in the
gluing and pre-tightening stage. The maximal travel of the
piezostack is 7\mum\ which translates into an unloaded movement of
about 80\mum\ at the fibre. This device thus allows us to stretch
the fibre by about one percent which is close to the maximum
elastic deformation tolerated by the silica glass and near the
value needed to tune one free spectral range.

\begin{figure}
  \begin{center}
  \psfig{figure=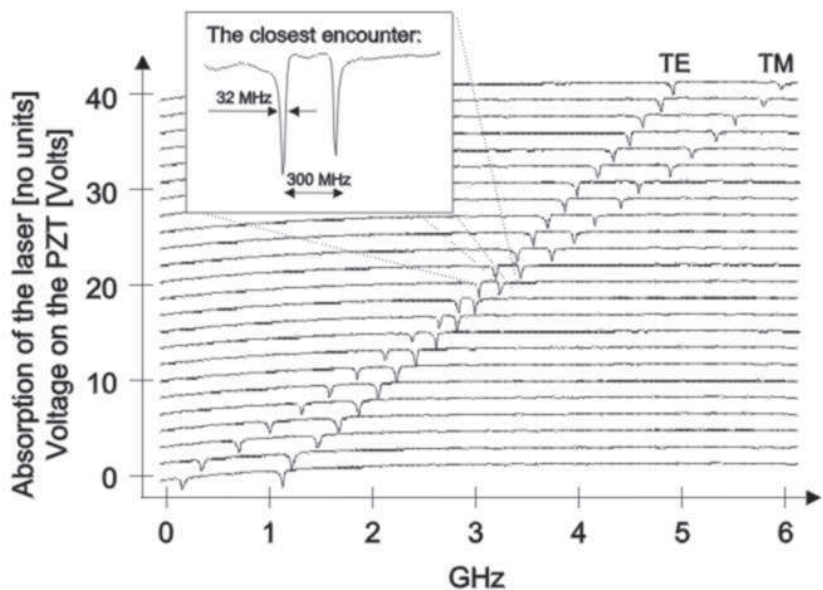,height=8cm} 
  \end{center}
  \caption{A series of scans of the WGM-resonances in a sphere against the
voltage applied on the PZT. The horizontal axis is the frequency
of diode laser. The vertical axis shows the transmission (in
arbitrary units) offset by the voltage of the PZT. A dip in the
transmission equates to an absorption by a whispering gallery
mode. For the sake of clearness the scans have been offset
proportionally to the voltage applied to the PZT. The insert shows
the closest point of the avoided crossing between the TE and TM
modes. The intrinsic quality factor of the sphere was
$Q=5\times10^{8}$.(see footnote \footnotemark[3])}
  \label{fig:tuning-v1}
\end{figure}

The average tuning range of the modes in the sphere has been
assessed by continuously increasing the voltage at the PZT and
observing the modes passing through the frequency window scanned
by the diode laser. Assuming there is no strong non linearity in
the tuning rate the average tuning range of the modes with respect
to the voltage applied to the PZT can thus be deduced. The
resulting maximal continuous tuning range is therefore 150\,GHz
which is about half of an FSR for the 210\mum\ sphere studied
here. A full FSR could not be reached mainly due to the stem being
significantly thinner than the sphere. Assuming a diameter of the
sphere twice as large as the one of the stem the stress on the
material will be four times as large on the stem. This leads to a
failure before the maximum tuning range for the sphere is reached.
The tuning range of 150\,GHz demonstrated here, however, already
suffices since it requires on average only two attempts to find a
sphere with predefined transverse and radial quantum numbers to
coincide with a given frequency, e.g. of an atomic line. An
analysis of the tuning shows that the major part of the tuning is
due to geometric deformation of the sphere. Therefore the modes
must increase in frequency with increasing strain, as can be seen
in figure~\ref{fig:tuning-v1}. It is therefore expected that the
slope of the tuning with respect to the voltage applied on the PZT
depends on the mode geometry as is apparent in
figure~\ref{fig:tuning-v1}. However, there is also a contribution
form the birefringence introduced by the stress: TM modes tune
more efficiently than TE modes. Figure \ref{fig:tuning-v1} shows a
series of scans of two resonances of opposite polarisation against
the voltage applied on the PZT. Clearly the TM mode tunes more
rapidly. For demonstration purposes the sphere was made from
inhomogeneous material. The quality factor of the sphere was
basically unaffected by this. \footnote[3]{In order to improve the
contrast in figure \ref{fig:tuning-v1} the gap between the sphere
and the prism was chosen to be very small. This increases the
coupling between the WGMs and the prism and thus enlarges the
modes in figure \ref{fig:tuning-v1} beyond their intrinsic line
width.} A $Q$ of $5\times10^8$ was measured. On the other hand the
coupling between certain modes of opposite polarisation was
greatly enhanced resulting in an avoided crossing of 300\,MHz (see
insert in figure \ref{fig:tuning-v1}). In `perfect' spheres the
coupling between modes is of the order of the coupling via
backscattering, i.e. a few hundred kilohertz.

\begin{figure}
  \begin{center}
  \psfig{figure=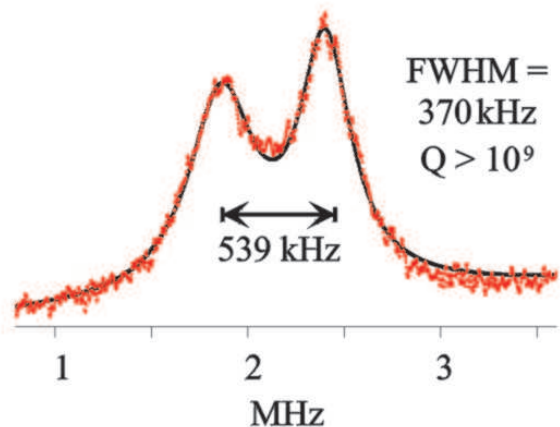,height=6cm}
  \end{center}
\caption{The intensity of the light absorbed by the sphere.  The
black line is a fit of two Lorentzian lines to the data. The lines
are 370\,kHz wide equating to a quality factor in excess of
$10^9$. The doublet originates in a coupling of
counter-propagating modes due to backscattering} \label{fig:Qv1}
\end{figure}

The long term stability of the modes is excellent. Over a number
of days the modes drifted less than 10\,GHz. On a shorter time
scale the stability was limited by fluctuations in the background
temperature. In \emph{all} the spheres produced by this method and
tested here we measured a quality factor of the order of $10^9$.
These are amongst the highest at this wavelength. Figure
\ref{fig:Qv1} shows one such measurement. The linewidth of
370\,kHz measured at 800\,nm corresponds to a $Q$ in excess of
$10^9$. The splitting of 539\,kHz between the peaks results from a
coupling of modes of opposite sense of rotation due to
back-scattering of the light by defects in the silica or residual
surface roughness \cite{WeissHaroche1995}. As expected, the
splitting remains constant even if the frequencies of the modes
are tuned. The quality factor can be preserved under atmospheric
conditions for a few days. A reduction in the $Q$ is usually
sudden and can often be traced back to a microscopic dust particle
settling in the vicinity of the WGM.

This first device can tune WGMs  of the microspheres of down to a
diameter of 60\mum\ by about half of an FSR whilst preserving very
good access and maintaining their high a quality factor of more
than $10^9$.


\subsection{The tuning device \#2}

\begin{figure}
  \begin{center}
  \psfig{figure=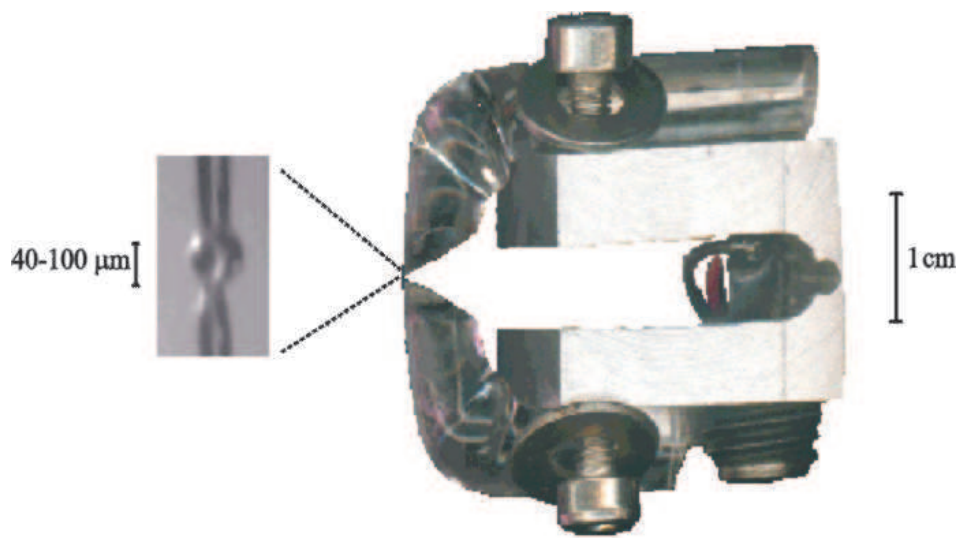,height=7cm}
  \end{center}
  \caption{The second tuning device. On the left of the device can
be seen an optical microscope image of the double stemmed
`sphere'}
  \label{fig:device-v2}
\end{figure}

For cavity quantum electronics experiments it is highly desirable
to use very small spheres in ultra high vacuum conditions. This is
difficult to realise with the first design due to the minimum size
of the spheres of about 60\mum. The second tuning device we
developed addresses this concerns.  It can be seen in
figure~\ref{fig:device-v2}. It consists of a U shaped base which
can be opened and closed with a screw and a vacuum compatible low
voltage PZT stack. Rods of pure silica are fixed onto the jaws of
the device and subsequently bent in an oxygen-propane flame to
meet at the centre in front of the device. The tips are then
ground to the shape of a pyramids with a tip to tip distance of
about 400\mum.

Next the $\mathrm{CO_2}$-laser is used to weld a short piece of
silica fibre across the gap between the tips of the pyramids. The
inevitable residual stress is removed by again gently heating the
material with the $\mathrm{CO_2}$-laser. The fibre is then placed
into the focus of two exactly counter-propagating laser beams
(lenses $f=25.4$\,cm). The material is carefully heated whilst the
tension on the PZT is continuously increased. This stretches the
material at the focus of the laser.  The procedure is repeated
some tens of micrometers below.  The result is a double neck in
the fibre.  The centre between the two indentations is then heated
strongly and the voltage slowly relaxed. The surface tension pulls
the material thus provided into a good approximation of a sphere.
(See insert in figure~\ref{fig:device-v2}) The relatively thick
stem assures that much of the deformation results in strain on the
sphere and does not just stretch the stems.  The indentations on
either side of the sphere reduce its ellipticity.

\begin{figure}
  \begin{center}
  \psfig{figure=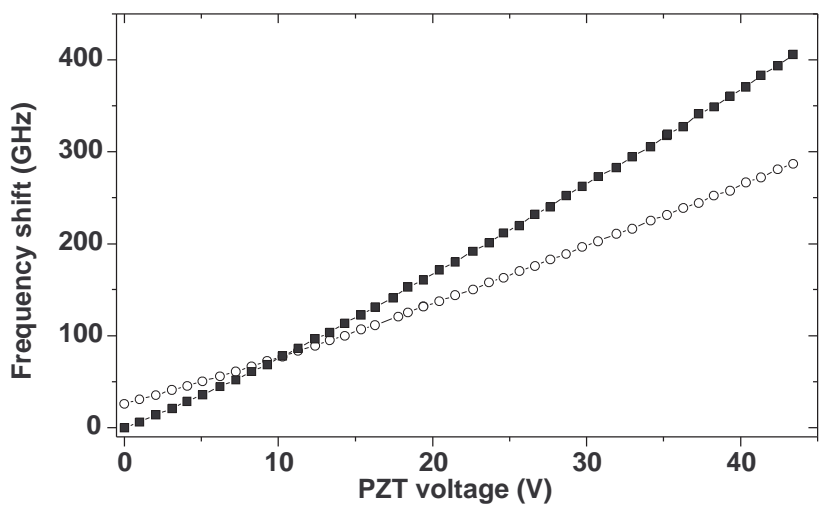,height=7cm}
  \end{center}
  \caption{Frequency shift for a TM~(\fullsquare) and a TE~($\circ$)
  mode followed continuously over the maximum tuning range. }
  \label{fig:tuning-v2}
\end{figure}

The second device was studied with a narrow linewidth tunable DBR
diode laser (Yokogawa YL78XNL). By scanning simultaneously the
laser current and the injection current into the Bragg grating
this laser can be tuned continuously by up to 1\,nm.  Its
linewidth is about 1\,MHz. The tuning of the WGMs could therefore
be observed directly over its maximum range.
Figure~\ref{fig:tuning-v2} shows the tuning of two WGMs in the
sphere against the voltage applied on the PZT. The frequency of
the resonance has been changed by 405\,GHz before the device
failed due to a fracture at the joint between the fibre and the
mount. Half of an FSR of the sphere has thus been scanned. Again,
as expected, the TE mode moved more slowly with the PZT voltage
than the TM mode.

The Q factor was lower than $10^9$, the value measured with device
1. This is probably due to some contamination of the sphere's
surface, possibly due to the deposition of a small amount of
crystallized silica. This will be avoided in the future by a
slight modification in the fabrication process.

The stability of the frequency of the modes is as mentioned of
particular concern to any future experiments with tunable
microspheres. The tuning was found to be perfectly reversible: On
the time scale of up to a week no drift of the modes, e.g. due to
plastic deformation of the sphere, could be observed.

The second device can tune the WGMs by about one half of an FSR.
Good access is still granted. Spheres of a diameter down to about
30\mum\ can be used.


\subsection{A comparison between the two devices}

The two different models of tuning devices presented above serve
quite different applications. The first one is clearly more
suitable for applications such as spectroscopy or spectral
filtering.  It is very simple to produce: A new resonator can
readily be made and coupled to a laser in less than one hour. It
might even be possible to mechanise such a production by adapting
well established pipette-pulling technology.  The tuning device
affords excellent access, good robustness, and ease of use. Its
main limitations are its minimum sphere diameter of 60\mum\ and
the relatively long stems necessary to fix the sphere to the jaws
of the device.

The second design has a potentially larger tuning range. It
permits smaller spheres to be used. Tuning of spheres down to less
than 40\mum\ has been demonstrated. It has excellent vacuum
compatibility and is more compact thus limiting the cost of vacuum
system. Its main drawback compared to the first device lies in the
more complicated production procedures requiring considerable more
skill and time. The second device is being used in ongoing CQED
experiments in our group.


\section {Conclusion}
A more detailed analysis has been reported of two novel devices
for microspheres which finally allows these extraordinary
resonators to be tuned into resonance with atomic and molecular
transitions. This opens the way toward a whole new range of
experiments in CQED and ultra sensitive spectroscopy.  It has now
become feasible to couple ultra cold atoms or quantum-dots to the
microspheres. The tunable microspheres will serve as affordable
'super' cavities for the detection of trace gases, e.g. by cavity
ring down spectroscopy.

\ack This work has benefited from the financial support of a CEE
TMR network contract ERBFMRXCT960066 "Microlasers and Cavity QED".

\newpage
\section* {Bibliography}
\bibliographystyle{Osa}
\bibliography{NJphys}
\end{document}